\documentclass[draft]{agujournal2019}
\usepackage{url} 
\usepackage{soul}
\draftfalse

\journalname{AGU Advances}

\begin{document}

%
%


\title{Reappraising the Elatina series: What if the Solar 
Interpretation Were Correct?}

%
%




\authors{F. Stefani\affil{1}}
\authors{M. Gugliotta\affil{2}}
\authors{G.M. Horstmann\affil{1}}
\authors{G. Mamatsashvili\affil{1,3}}
\authors{T. Weier\affil{1}}


\affiliation{1}{Helmholtz-Zentrum Dresden-Rossendorf, Bautzner Landstr. 400, 01328 Dresden, Germany}
\affiliation{2}{Institute of Geosciences and Geography, 
Martin‐Luther University Halle‐Wittenberg, 06120 Halle (Saale), Germany}
\affiliation{3}{Abastumani Astrophysical Observatory, Mount Kanobili, Abastumani 0301, Georgia} 





\correspondingauthor{F. Stefani}{F.Stefani@hzdr.de}



\begin{keypoints}
\item The Elatina series is reconsidered with view on its explainability as varves related to solar activity.
\item The negative correlation between cycle duration and amplitudes
of subsequent cycles, and an 90$^{\circ}$ phase jump, are in line with the solar 
theory.
\item The refined cycle durations are used to infer ancient planetary orbits.
\end{keypoints}

%
%

%
%


\begin{abstract}
We revisit the sedimentary laminae from the Neoproterozoic 
Elatina Formation of South Australia. These were first interpreted 
as varves driven by solar activity and later as tidalites formed by 
tidal processes, although the debate has remained open. This series 
exhibits remarkably consistent periodicities, including a primary 
and a secondary period of approximately 12 and 314 laminae, respectively.
Our present analysis reveals a statistically significant negative 
correlation between the length of a cycle and the amplitude of the subsequent cycle.
By analyzing the residuals of the series' minima with respect to a 
linear trend and calculating Dicke's ratio, we show that this series 
exhibits a high degree of phase stability, except for a single 
break-point which possibly indicates a  90$^{\circ}$ phase shift.
As these findings are difficult to reconcile with the tidal model, 
we reconsider the original interpretation as solar-driven 
sedimentation and tentatively discuss it in terms of a recently 
developed synchronization model of the solar dynamo. This model 
is used to calculate the orbital periods of Venus, Earth, Jupiter 
and Saturn that would be required to explain the 12-year cycle, 
when interpreted as a modified solar Schwabe cycle, and the 314-year cycle, 
when interpreted as a prolonged Suess-de Vries cycle. 
Assuming that the sums of the angular momenta of 
Jupiter and Saturn and Venus and Earth are pairwise conserved, we find 
that the planetary orbits change surprisingly little.
The plausibility of such changes 
over a period of seven hundred million years 
is discussed in light of solar system dynamics.
\end{abstract}

\section*{Plain Language Summary} 
The sedimentary laminae of the Elatina Formation in South Australia exhibits 
two stable periodicities: a primary period of around 12 laminae and a 
secondary period of 314 laminae.
While these laminae were originally interpreted as yearly varves which 
reflect changes in solar activity and climate, this interpretation has 
recently shifted towards the tidal theory. According to this theory, the 
laminae correspond to subdaily sedimentation.
We re-examine some statistical properties of the Elatina series and argue 
that they can be more easily explained in terms of the original solar 
theory. Based on this finding, we attempt to infer the ancient orbital 
periods of Venus, Earth, Jupiter and Saturn from the 
observed periodicities. To do so, we rely on a recently developed 
synchronization model of the solar dynamo.

%
%

%


%
%
%
%

\section{Introduction}

Over four decades ago, the presence of alternations of light- and dark-coloured laminae 
of the Neoproterozoic Elatina Formation at Pichi Richi Pass, in South Australia, were 
reported first in outcrops and sawcuts \cite{Williams1981}
and then from drillcores \cite{Williams1985a}. The exact 
age of this unit remains uncertain, as there is no direct age dating; nonetheless, 
more recent publications place it at around 635 Ma 
(i.e. at the boundary between the Cryogenian and Ediacarian) based on correlations 
to other units worldwide considered as part of the 
same glaciation \cite{LeHeron2012,Williams2019}. The distribution of these laminae 
and their thicknesses recorded cyclic patterns, which were interpreted in these 
earlier studies and successive ones \cite{Bracewell1985,Williams1985b,Bracewell1986} 
as varves forming in a prodelta setting of a broader periglacial system and 
actually preserving evidence of solar-driven climatic cycles. Although the 
entire 9.39\,m long drillcore entailed a series of approximately 19,000 laminae, 
most of the detailed studies was drawn from a sequence of 1,337 laminae obtained 
from enlarged photographs of thin sections. When interpreted as a proxy for 
annual solar activity, the exceptional clarity of this series is striking, 
making it comparable, or even superior, to more recent sequences of cosmogenic 
isotopes \cite{Brehm2021,Usoskin2025}.  A first suggestion that these cycles 
could have been tidal in origin came from \citeA{Goguel1982}, but was firmly 
rejected by \citeA{Williams1982} who insisted 
on the periglacial origin with laminae accumulating in deeper water due to 
density flows. Nonetheless, a few years later, the view that these laminae 
and their cycles are tidal in origin, although still in the context of a 
periglacial setting, was supported by Williams and coauthors
\cite{Sonett1988,Williams1988,Williams1989}. The tidal interpretation became 
gradually more popular and is the one more often proposed up to recent 
times \cite{LeHeron2011,LeHeron2012,Williams2019,Williams2023}. In the case 
of the tidal view, the laminae would not represent yearly records, but the 
ones of tidal cycles, as for example, neap-spring cycles occurring 
approximately every fortnight \cite{Zahnle1987,Deubner1990,Reineck1990,Mazumder2005}. 
Nonetheless, the alternative view as varves and solar cycles has not 
been entirely abandoned and was still proposed recently \cite{Tyasto2020}, 
maintaining the two theories and the related debate open.

In this context, the problem of how to comprehend the alleged nexus 
between solar activity and climate was raised by some authors.
In his early papers, Williams tried to explain this phenomenon by 
suggesting that the low level of oxygen in the Precambrian 
atmosphere would have resulted in a much lower ozone layer, 
allowing ultraviolet (UV) radiation to penetrate deeper 
into the atmosphere  \cite{Williams1985a}.  Changes in local 
ozone levels, triggered by variations in solar UV radiation, 
may have had a direct impact on ground-level temperatures.
Actually, this idea is not incompatible with other possible 
mechanisms, such as the top-bottom influence of UV variations 
on cyclone paths, as proposed by \citeA{Prasad2004} to explain 
the impact of solar activity on sediment layers in the 
Lake Lisan region in recent times.

Presupposing its solar origin, Bracewell's decisive idea
for the quantitave analyisis of the Elatina series 
was to de-rectify the $\sim$12\,-year 
Schwabe-type cycle to a $\sim$24-year Hale-type cycle 
\cite{Bracewell1986,Bracewell1988a,Bracewell1988b}.
Using this method, the author was able 
to constrain the main Hale cycle to 23.7\,years, 
with a very low 
standard deviation of just 0.2\,years. This remarkable 
precision of the ``Elatina oscillator'', and 
the resulting high Q-factor
of 120, led him to argue in favor of 
some clock mechanism behind solar activity. 
Interestingly, while citing Dicke's 
work \cite{Dicke1978} on the distinction between 
random walk and clocked processes, Bracewell 
did not analyze the Elatina series 
from this perspective.

Our motivation for taking up the ``sunny trail'', 
as originally paved by Williams, Sonett and Bracewell, 
stems from our interest in this potential  ``clock'' of the 
solar dynamo. The underlying synchronization theory
(which traces back to pioneering work of 
\citeA{Hung2007,Scafetta2012,Wilson2013,Okhlopkov2016})
relies on the tidal action of the 
orbiting planets on the Sun (not to be confused with
the tidal theory
of the Elatina series which considers the effect of Moon and Sun on the Earth).
Following the 
application of simple ODE and 1D-PDE solar dynamo models 
\cite{Stefani2016,Stefani2018,Stefani2019,Stefani2021}, 
the synchronization theory has matured significantly
in recent years  \cite{Stefani2023,Klevs2023,Horstmann2023,Stefani2024}.
An account of its history, including 
the debates with opponents, 
can be found in 
\citeA{Stefani2025}.

In short, this model builds 
on the premise that
magneto-Rossby waves at the solar tachocline
\cite{Dikpati2017,Dikpati2020,Raphaldini2015,Raphaldini2019,Zaqarashvili2010,Zaqarashvili2021} 
form a perfect resonance ground 
for the three
spring tides with periods of 118, 199 and
292 days, as exerted by the pairs of the 
tidally dominant planets Venus, Earth and Jupiter
\cite{Horstmann2023,Stefani2024}. While the dominant beat 
of these three waves has a period of  
1.723 years \cite{Stefani2025,Stefani2026}, which is 
well-known in solar 
physics as the quasi-biennial oscillation (QBO),
a secondary
beat period is 11.07\,years, aligning closely
with the average Schwabe cycle. A longer-term 
Suess-de Vries-type cycle of 193\,years arises in this model 
as a secondary beat between the 22.14-year Hale 
cycle and the period of the rosette-shaped motion of the Sun 
around the solar system's barycenter.
While this motion is known to be dominated by the 19.86-year cycle 
of Jupiter and Saturn, the precise mechanism by which any dynamo-relevant 
internal motion can result from it is still under scrutiny 
\cite{Wilson2013,Solheim2013,Shirley2006,Shirley2023,Pierron2026}.

With $P_V$, $P_E$, $P_J$, $P_S$, $P_{Sch}$, $P_{Hal}$, $P_{Bar}$, $P_{SdV}$ 
denoting, respectively, the orbital periods of Venus, Earth, Jupiter, and Saturn, 
the Schwabe and Hale cycles, and the barycentric and Suess-de Vries cycles, 
the three governing equations of the synchronization model 
are
\begin{eqnarray}
P_{Hal}&=&2 P_{Sch}=\left(\frac{3}{P_V}-\frac{5}{P_E}+\frac{2}{P_J} \right)^{-1}  \; , 
\end{eqnarray}
\begin{eqnarray}
P_{Bar}&=&\frac{P_J P_S}{P_S-P_J}  \; , 
\end{eqnarray}
and
\begin{eqnarray}
P_{SdV}&=&\frac{P_{Hal} P_{Bar}}{P_{Hal}-P_{Bar}}  \;.
\end{eqnarray}

Quite reassuring for the validity of this model is
the remarkable coincidence (see Figure 9 of \citeA{Stefani2024}) 
of the spectral peaks resulting from it
with those of climate-related sediment data from Lake Lisan 
\cite{Prasad2004}. 
Also relevant is the high degree of self-consistency of the 
model, whose explanatory skill 
ranges now from Rieger-type periods and the QBO (for the quite accurate 
agreement with observations, see \citeA{Stefani2026}) to the 
Schwabe/Hale cycle and finally the Gleissberg and Suess-de Vries cycles.

Given this solar-physics background, we thought it might 
be worth trying to apply the theory to the Elatina series.
When tentatively accepting the double-synchronized solar dynamo 
as the mediator, the planets' orbits must also change in order to 
account for the observed variations in period.
The only problem is: by how much?
Furthermore, how consistent are the derived orbital changes with 
solar system dynamics over a period of more than 600 million years?

To make the paper self-contained, we begin by illustrating the 
main data from the Elatina series. 
Then, with a view to corresponding relations for the current solar dynamo, 
we will investigate various correlations between cycle length and amplitude.
We will also analyze phase stability and clocking in terms of Dicke's ratio 
 and discuss the possibility of the occurrence of 90$^{\circ}$ phase jumps. 

Encouraged by the 
similarities with the present solar dynamo
(and the concurrent difficulty of explaining the observations 
in terms of the tidal theory), we will then employ Equations (1-3) 
to infer the planetary orbits at the time of the formation 
of the Elatina series. 
As will be demonstrated, only 
minor changes to the angular momentum of the planets 
(around 1 per cent for Jupiter and 0.005 per cent for Earth) 
are required to account for the significant changes in the 
two observed periods (7 per cent for Hale and 63 per cent 
for Suess-de Vries).
In the discussion we will come back to the 
caveat that the Elatina series might still be 
interpreted in terms of ebb-tides deposits.

\section{The Elatina series and its statistical properties}

In absence of the original data set we have
digitized, similarly as
\citeA{Tyasto2020}, the Elatina series data as published 
by \citeA{Bracewell1986}.
Figure 1a, which corresponds mainly to Figure 
1 of that paper,
shows the series of measured thicknesses 
of 1337 laminae which we will tentatively 
interpret as varve years.

\begin{figure*}
\centering
	\includegraphics[width=0.8\textwidth]{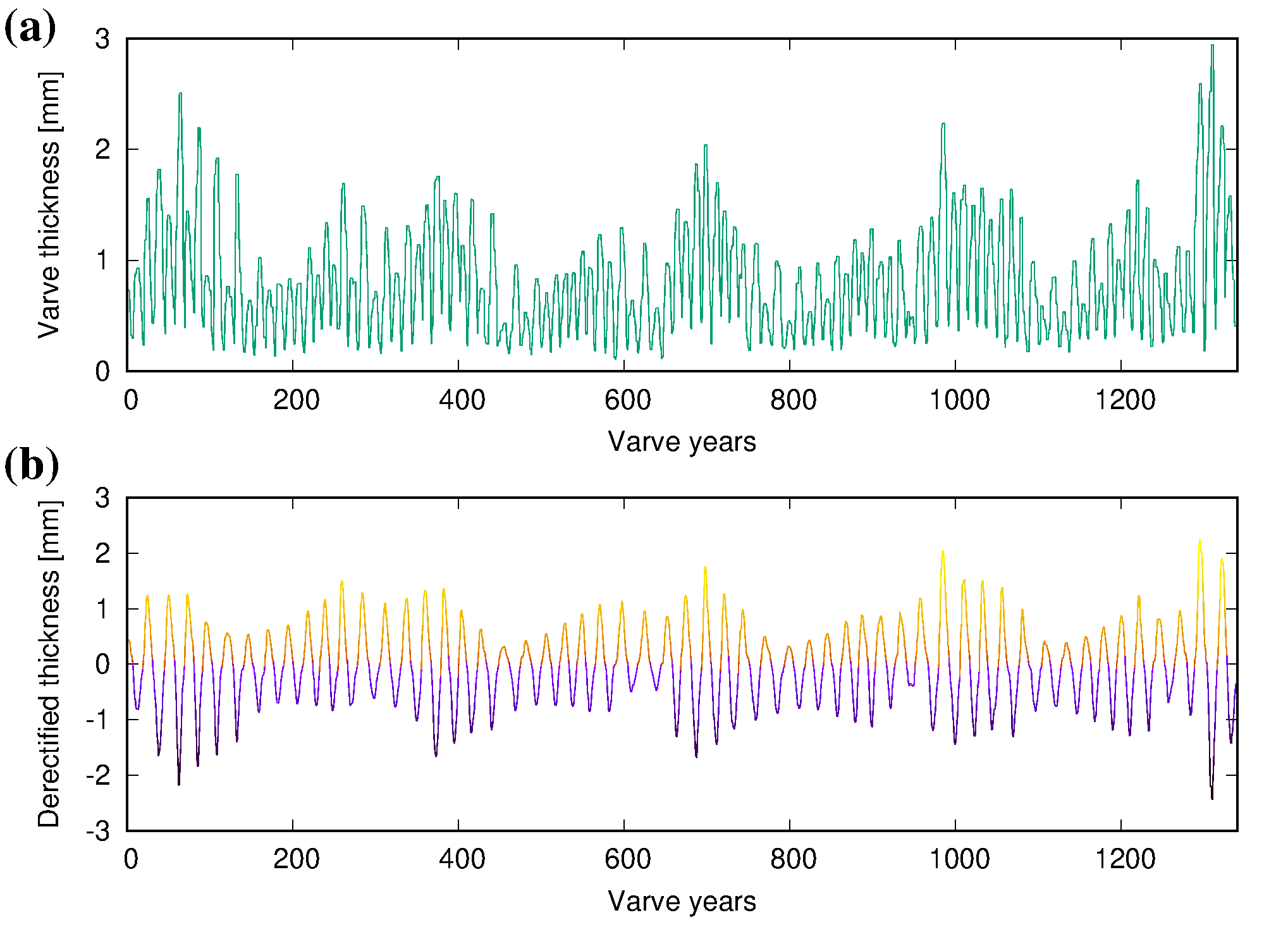}
	\caption{(a) Series of laminae thicknesses in mm, as 
    digitized from
    Figure 1 of \citeA{Bracewell1986}. (b) De-rectified 
    alternating series, as digitized from 
    Figure 2 of \citeA{Bracewell1986}.}
	\label{fig:fig1}
\end{figure*}

Even better digitization results can be 
obtained from the de-rectified series of Figure 
2 of \citeA{Bracewell1986} which is replotted here
in Figure 1b. Since we are mainly interested 
in the minima and maxima, we consider
the accuracy of our digitizations to be sufficient.
As for the values of the modified ``Hale cycle'' we will,
at any rate, rely on the value 23.7\,years as previously 
derived by \citeA{Bracewell1988a}.
We will also use his value of 314\,years for the 
Elatina cycle, which will be identified with a
modified Suess-de Vries cycle, whose present 
period is approximately 200 years (or 193 years,
according to the synchronization theory).

\begin{figure*}
\centering
	\includegraphics[width=0.99\textwidth]{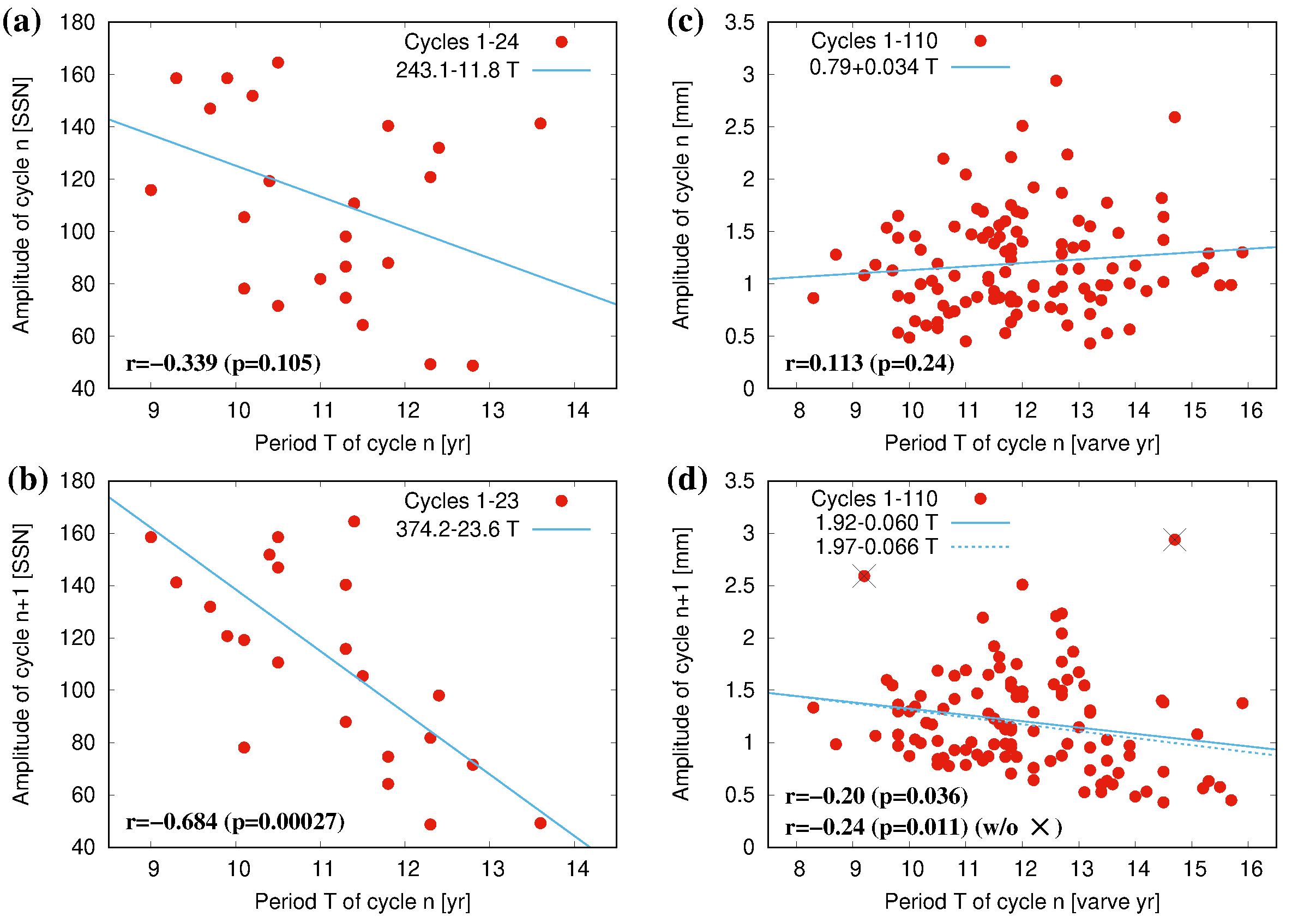}
	\caption{Relation between cycle length and amplitude for the 
    present solar dynamo and the Elatina series.
    (a) Amplitude of cycle $n$ over the length of cycle $n$ for the most recent
    24 solar
    cycles. The emerging correlation of -0.339 is not significant.
    (b) Same for the amplitude of cycle $n+1$. The correlation of 0.684 is highly significant.
    (c) Amplitude of cycle $n$ over the 
    length of cycle $n$ for the
    110 cycles of the Elatina series. 
    The correlation of 0.113 is not significant.
    (d) Same for the amplitude of cycle $n+1$. The correlation of 0.2 
    is significant on the 95 percent level. This would still improve 
    if the two very strong amplitudes, marked by the crosses, are omitted. }
	\label{fig:fig1}
\end{figure*}

The first analysis we perform on these data is to determine whether there 
is a relationship between cycle length and amplitude. Figures 2a and 2b 
show the corresponding relations for the present solar dynamo. Using the 
cycle maxima determined by the 13-month mean of the International 
Sunspot Numbers, as taken from Table 1 of \citeA{Hathaway2015}, in 
Figure 2a we plot the cycle amplitude over the cycle length.
The resulting (Pearson)
correlation coefficient turns out to be -0.339 which corresponds 
(for the 24 cycles considered) to a $p$-value of 
0.105, which is not significant. A much stronger 
correlation of -0.684, with a very low $p$-value of 
0.00027, emerges only when we consider
the dependence of the amplitude of 
the  $(n+1)^{th}$ cycle on the 
length of the $n^{th}$ cycle, as shown in Figure 2b.

Panels (c) and (d) of Figure 2 show the same for the 110 
``solar cycles'' 
as they would result from
the Elatina series. Here, the cycle period was inferred from the 
differences of two subsequent zero-crossings of the de-rectified
series shown in Figure 1b. For the correlation of the $n^{th}$ cycle 
length
with the $n^{th}$ cycle amplitude, Figure 2c shows a slightly 
positive value of 0.113, which is not significant 
($p$-value 0.24). More interestingly, the correlation between 
the $n^{th}$  cycle 
length with the $(n+1)^{th}$ cycle amplitude acquires a value 
of -0.2 whose $p$-value of 0.036 indicates already
a significance on the 95 percent level. If we had rejected 
the two extreme values (marked by the crosses in Figure 2d), 
the correlation would have become -0.24. With the resulting
$p=0.011$ we would almost reach
significance at the 99 percent level.

Although it does not provide irrefutable evidence 
for a solar origin of the Elatina series, we find the 
qualitative agreement with the current solar dynamo quite remarkable.
In contrast to the well-established theoretical basis for 
the link 
between the duration of a solar cycle and the strength of the subsequent 
one, we are not aware of any potential explanation in frame of the
the alternative tidal theory.

In the following, we will further assess the sequence of minima and maxima 
of the Elatina series, focusing particularly on phase stability and 
clocking.
The first step in this process is to determine the residuals
$\delta_n=t_n - n \tilde{P}_{Sch}$, i.e. the distances 
of the $n$-th
instant $t_n$ of a minimum (or maximum) 
from the corresponding instant that 
would follow from a purely
linear trend when choosing  $\tilde{P}_{Sch}$ 
as the Schwabe-cycle period. 
Using the cycle minima, those residuals 
are shown as the violet curve 
in Figure 3a, where we have chosen
$\tilde{P}_{Sch}=11.95$\,years, 
which is slightly longer than the
11.85\,years as advocated by
\citeA{Bracewell1986}. Indeed, our choice leads to 
a relatively horizontal appearance of the 
curve, at least piecewise,
while a certain jumpy behavior becomes 
visible around the year 600.

To assess phase stability, we employ Dicke's ratio 
\begin{eqnarray}
D&=&\frac{ \sum_{i=2}^{N} \delta^2_i}{\sum_{i=2}^{N} (\delta_i-\delta_{i-1})^2}
\end{eqnarray}
between the mean square of the residuals 
$\delta_i$ and the mean square of the
{\it differences} $\delta_i-\delta_{i-1}$ 
between two subsequent residuals. As 
shown by \citeA{Dicke1978}, the dependence of this ratio on 
the number $N$ of considered cycles 
reads $(N+1)(N^2-1)/(3(5 N^2+6N-3))$ for a random walk 
process, and $(N^2-1)/(2(N^2+2N+3))$
for a clocked process. Hence,
for a random-walk process, $D$ 
converges for $N \rightarrow \infty$ towards 
$N/15$, while for a clocked process it converges 
towards the constant value of 0.5.   
Figure 3b shows both theoretical curves, together 
with Dicke's ratio computed for the actual residuals of the 
cycle minima.      
Note that the number $N$, 
shown on the  upper abscissa, 
increases when moving from right to left, i.e. when decreasing 
the starting year from the terminal year 1337.

Two distinct features become immediately visible in the violet curve of Figure 3b.
Starting close to 
the right end (i.e., for low $N$), the curve first 
clings to the (gold) clocked-process curve, but then jumps up 
around the year 1200, quickly swinging above 
the (light blue) random-walk curve. This overshooting is typical of 
processes governed by long-term periodicities (see Figure 2 in 
\citeA{Stefani2019}). 
Moving further to the left, the curve remains horizontal, falling again below 
the linearly increasing random-walk curve. However, around the year 
600, it suddenly jumps up again before continuing as a horizontal curve.

\begin{figure}
	\includegraphics[width=0.99\textwidth]{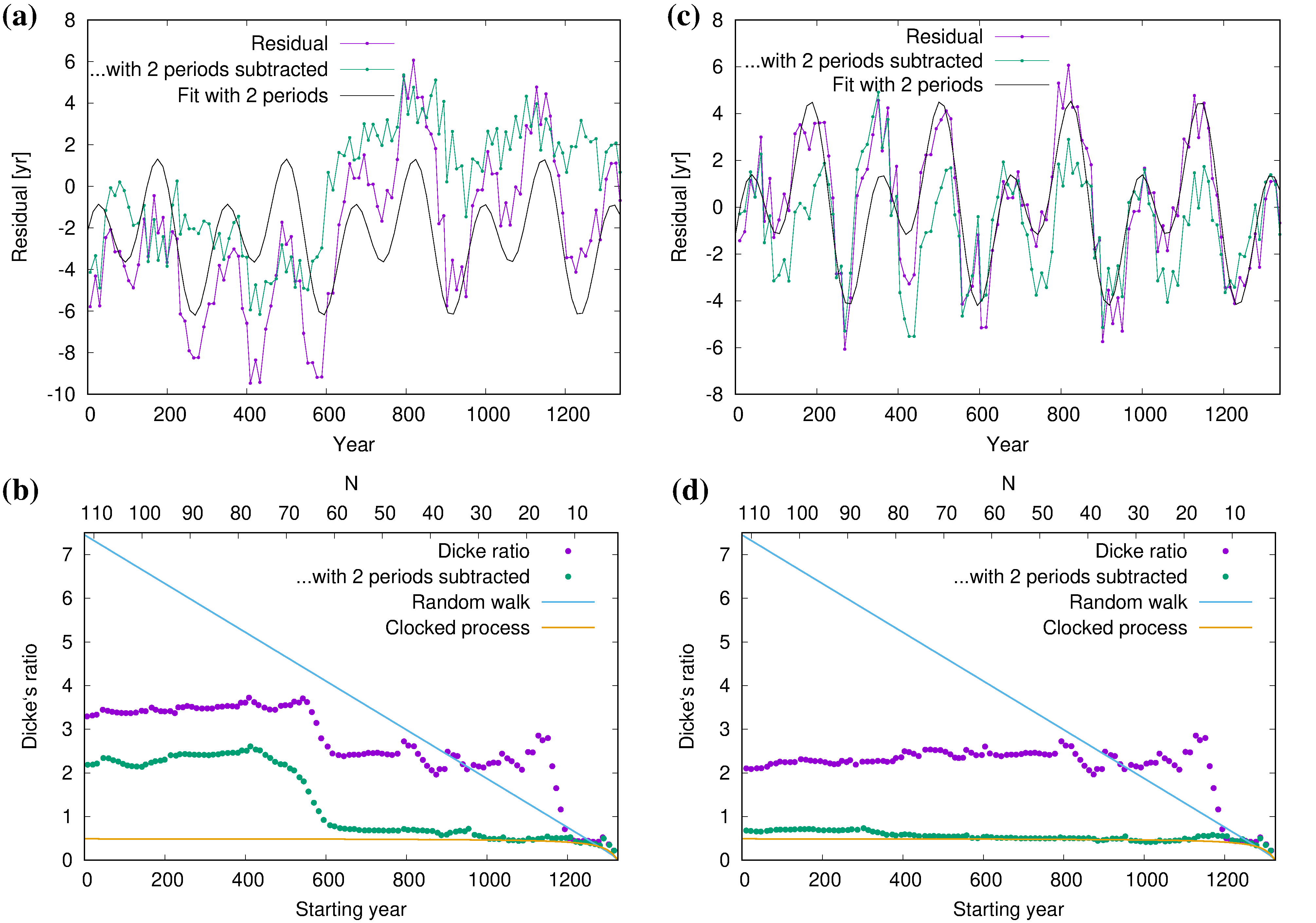}
	\caption{Residuals of the minima of the 
    Elatina series, and Dicke's ratio derived from them. 
    (a) Residuals (violet) of the minima of the 
    Elatina series, as inferred from Figure 1a, when assuming 
    $\tilde{P}_{Sch}=11.95$\,years.
    The optimal sine-fit (black) of these residuals 
    with an  Elatina-type period $\tilde{P}_{Ela}$,
    and its first overtone, yields 
    $\tilde{P}_{Ela}=322.9$\,years, which is remarkably 
    close to the 314 years as inferred 
    {\it{from the amplitudes}} by \citeA{Bracewell1988a}.
    The green curve shows the residuals with
    this two-period fit (black) subtracted from 
    the original (violet) residuals.
    (b) Dicke's ratio for a random-walk (light blue), 
    a clocked process (gold), and for the two curves from (a),
    as a function of the starting year which also 
    determines
    the number $N$ of cycles taken into account (shown 
    on the upper abscissa).
    Dicke's ratio for the original residuals (violet) 
    shows a jump around the starting year 1200 which is
    a typical feature for processes with underlying dominant
    long-term periods. After subtracting these long-term
    periods, this jump disappears (green). (c) Same as (a), 
    but with the minima 
    replaced by the maxima before the year 600. 
    Here we use again $\tilde{P}_{Sch}=11.95$\,years, and 
    obtain $\tilde{P}_{Ela}=319.2$\,years 
    from the double-period fit. (d) Same as (b), but with the minima 
    replaced by the maxima before the year 600. Note 
    the complete disappearance of the jump close to the year 600.}
	\label{fig:fig2}
\end{figure}

To isolate the different effects, we subtract the 
long-term periods from the residuals' curve shown 
in Figure 3a. Even visually, these periods appear 
similar to the so-called ``Elatina cycle` of 314 years 
\cite{Bracewell1986} and its first overtone.
The black curve in Figure 3a represents an
optimal fit of the violet curve 
with two sine-functions whose periods 
turn out to be 323 and 161 year, respectively.
This result is by no means self-evident. 
While the Elatina cycle of 314 years was originally 
inferred from amplitude variations, a similar value is 
obtained here from {\it variations of the residuals}.
Note that a similar correspondence was previously found for the
present-day Suess-de Vries cycle, see 
Figure 2 in \citeA{Stefani2020}.
Clearly, the anti-correlation between the cycle period
and amplitude, as discussed above, 
provides the most straightforward explanation for 
this correspondence.

After subtracting from the original residuals 
(violet) the optimal double-sine fit (black), we arrive at the 
green curve in Figure 3a. Overall, this curve generally shows less variance, 
while the jump close to the year 600 becomes even more pronounced.
Remarkably, the corresponding (green) curve for Dicke's ratio in Figure 3b
now closely follows the clocked process, at least down to the starting 
year 600. Then it undergoes a significant jump, suggesting a break 
in the process at this point.

In a first attempt to explain this jumpy behavior,
we considered the possibility that
one cycle somewhere 
between the years 588-658 might 
have been missed (indeed, there are five 
consecutive cycles with a remarkably long 
average duration of 14 years).
To assess this possibility, we first
``smuggled in'' one putative additional minimum 
at the year 624. Although slightly reduced, the
jump in Dicke's ratio around the year 600 remained.
Evidently, just inserting 
one additional minimum, i.e. by 
introducing a phase jump by 180$^{\circ}$, 
does not solve the problem.

But what about a phase jump of 90$^{\circ}$?
In the context of solar dynamo theory, this option might 
sound very strange. For a clocked, but noisy, dynamo process one could 
easily understand some irregular breakdowns, after which 
the normal cycle returns with a 180$^{\circ}$ phase shift.
By contrast, 90$^{\circ}$ phase jumps have only been 
discussed so far by \citeA{Vos2004}  
in their analysis of algae-related proxies of the solar 
cycle in the early Holocene. However, the authors 
explained these 90$^{\circ}$ phase jumps in terms of 
the optimal growth conditions of the investigated 
algae rather than in terms of the underlying solar cycle.

Therefore, we were very surprised when we found 
90$^{\circ}$ phase jumps in numerical simulations of 
a refined synchronization model of the solar dynamo \cite{Stefani2025}.
In this model, in which the 1.723-year QBO plays a key role,
the 11.07-year period only appears as an envelope in 
the $\alpha$-effect, the maximum of which may 
shift by up to 5.5\,years when the time average changes.
Apart from the algae-data from the Holocene \cite{Vos2004}, 
another candidate for an 90$^{\circ}$ phase jumps 
appears around 1795 A.D. (see Figure 2 in \citeA{Stefani2019}).
A more detailed analysis of this time, for which 
 \citeA{Usoskin2001} discussed the problem  of a ``missing cycle'' ,
must be left for future work.

Coming back to the Elatina series: in 
a first attempt to assess the possibility of a 
90$^{\circ}$ phase-jump effect, we 
replaced the minima of the series by 
the maxima before the year 600.
The results of this rather ``bold'' 
procedure are shown in Figures 3d. 
Remarkably, the former jump of Dicke's ratio close 
to the year 600 disappears entirely, as if a 90$^{\circ}$ 
phase shift at this point perfectly described the process.
We will return to this point in the conclusion, where we 
will discuss its implications for choosing between the 
solar and tidal theories of the Elatina series.

For the moment we just conclude that the Elatina series
shows a very strong sign of phase-stability and clocking,
just with one phase jump around the year 600, which would be hard to
reconcile with the tidal theory. The anti-correlation between
the duration of a cycle and the amplitude of the 
following cycle speaks also in favour of the solar theory.

\section{Inferring ancient orbital periods}

Having obtained decent arguments in 
favour of the solar interpretation of the Elatina series, 
 we now ask what changes of the planetary orbital period 
 would be required to accommodate the modified Hale and 
 Suess-de Vries cycles as inferred from the series. 
While it is not our goal to delve into the challenges 
of planetary system dynamics for several hundred million years 
(see, for example, \citeA{Sussman1992} and \citeA{Laskar2008}), we will 
occasionally take a brief look at some known results in order 
to assess the plausibility of our own findings.

The starting point is the expression for the angular momentum
 $L_i$ of a planet with
an orbital period $P_i$ and an eccentricity $e_i$, 
under the simplifying assumptions of negligible inclination 
and a mass $m_i$ 
that is very small compared to the mass $M$ 
of the central star:
\begin{eqnarray}
L_i=m_i (GM/\sqrt{2 \pi})^{2/3} P^{1/3}_i \sqrt{1-e^2_i} \;.
\end{eqnarray}
Evidently, the angular momentum depends on both the
orbital period  $P_i$ and the eccentricity $e_i$. 
As mentioned, accurately determining their joint time evolution 
under the mutual influence of the other planets is a highly 
ambitious task that will not be considered here. However, 
we note that according to the Laplace–Lagrange secular 
evolution theory the eccentricity of Jupiter is 
restricted to values between 0.0256 and 0.0611, 
while that of Saturn is restricted to values
between 0.0121 and 0.0845 
(see Table 10.4 in \citeA{Fitzpatrick2012}).
Typically, both eccentricities evolve 
in opposite directions \cite{Michtchenko2001} in order
to fulfill the conservation of the total angular 
momentum. According to Equation (5), for Jupiter this 
would amount to an angular momentum variation of 
0.15 per cent.

\subsection{Assuming unchanged eccentricity}

In order to gain an initial understanding of the typical changes in 
orbital periods and angular momenta, we will first consider the 
case of unchanged eccentricities, which we acknowledge is not very realistic.
Modified eccentricities will be considered further below.

Specifically, we assume  a period of 23.7 years 
for the modified Hale cycle, and identify the 314-year Elatina period 
with a modified Suess-de Vries cycle.
While the former identification is obvious given the relatively minor 
change of seven per cent, the latter is certainly contentious.
It is indeed reasonable to have serious doubts about the idea of attributing 
the 63 per cent increase in the 193-year Suess-de Vries cycle to very minor 
changes in orbital periods.

However, we should bear in mind that the beat periods in Equations (1) 
and (3) are highly sensitive to slight differences between large 
numbers in the respective denominators.  The following demonstrates 
that only minor adjustments to the planetary orbits are required to 
account for significant 
changes to both the Hale and Suess-de Vries cycles.

In the first instance, our inverse problem of inferring the four 
periods and eccentricities of Venus, Earth, Jupiter and Saturn from 
only two modified Hale and Suess-de Vries cycles is clearly ill-posed. To make the problem
tractable, we split it into two parts.
First, we solve Equation (3) to express $P_{Bar}$ 
in terms of $P_{Hal}=23.7$\,years and $P_{SdV}=314$\,years, 
leading to
\begin{eqnarray}
P_{Bar}&=&\frac{P_{Hal} P_{SdV}}{P_{Hal}+P_{SdV}}=22.04\,{\mbox{years}} \;.
\end{eqnarray}

Having derived $P_{Bar}$, we now use Equation (2) 
to obtain $P_J$ and $P_S$. Yet, this is only possible
when taking into account a second equation, for which 
we use the conservation of the sum of the angular momenta of
Jupiter and Saturn. 
Assuming that the two eccentricities 
remain unchanged from their current values, we obtain from Equation (5) 
the relation
\begin{eqnarray}
P_{S}&=&\left( \frac{m_J \sqrt{1-e^2_{J,t}}}{m_S \sqrt{1-\epsilon^2_{S,t}}} (P^{1/3}_{J,t}-P^{1/3}_{J})+P^{1/3}_{S,t} \right)^3  \;,
\end{eqnarray}
where $P_{J,t}=11.86$\,years, $e_{J,t}=0.049$, 
$P_{S,t}=29.46$\,years,
$e_{S,t}=0.054$ denote the periods and 
eccentricities 
of Jupiter and Saturn as of {\it today}.
Then, in Figure 4a,  we solve Equation (2) graphically 
using this two-planet angular momentum constraint 
of Equation (7).
From the intersection point of the 
green and the black curves, we read off
$P_{J}=12.21$\,years and,
correspondingly, $P_{S}=27.38$\,years.
For Jupiter this amounts to an increase of $P_J$ by
2.95 per cent while for Saturn $P_S$ decreases by 
7.06 per cent. In view of the 1/3 power in 
Equation (5), these values correspond
to angular momentum changes of 
0.98 and 2.35 per cent, respectively.

\begin{figure}
	\includegraphics[width=0.99\textwidth]{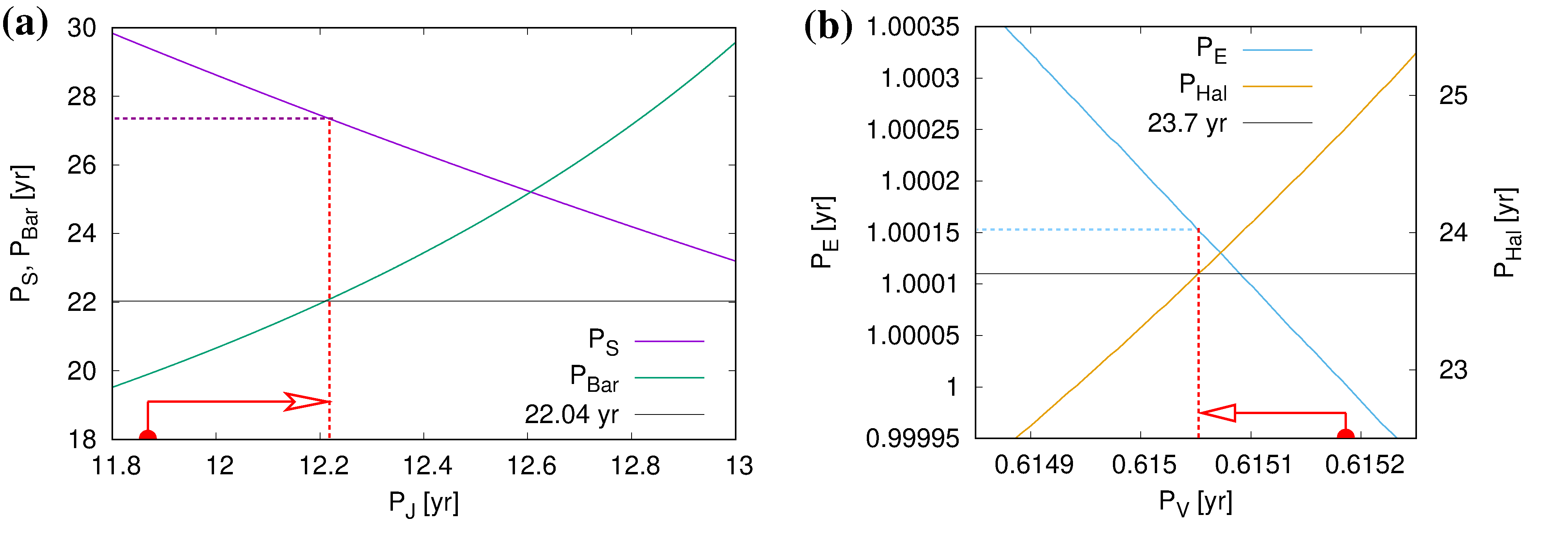}
	\caption{Graphical solution of Equation (2). (a) Assuming  
    the pairwise angular-momentum constraint of Equation (7) for 
    Jupiter and Saturn. The period of Jupiter must be shifted
    (red arrow)
    from 11.86 to 12.21 years, and that of Saturn from 
    29.46 to 27.38 years, so that the arising period 
    of the 
    motion around the barycenter becomes 22.04 years, as derived 
    in Equation (6). (b) Assuming  the
    pairwise angular-momentum constraint of Equation (8) for 
    Venus and Earth. When utilizing $P_J=12.21$\,years, as 
    derived in (a), the period of Venus must be shifted
    (red arrow)
    from 0.61519 to 0.61505 years, and that of Earth from 
    1.0 to 1.00015 years, so that the 
    arising period of the Hale cycle becomes 23.7 years as inferred 
    from the Elatina data.
    }
	\label{fig:fig5}
\end{figure}

Before discussing the plausibility of such values in terms of celestial 
mechanics any further, we first complete our solution by determining 
the corresponding period changes for Earth and Venus. For this 
purpose, we re-enter Equation (1) with the modified Jupiter 
period $P_{J}=12.21$\,years
and  $P_{Hal}=23.7$\,years.
Once again, we require a second equation in order to determine both $P_V$ and $P_E$.
In contrast to the physically well-justified 
pairwise angular momentum 
constraint for the two most massive 
planets Jupiter and Saturn,  the application of this 
principle to Venus and Earth is
less plausible. If we apply it still, 
and also keep the present eccentricities
$e_{V,t}=0.007$ and $e_{E,t}=0.017$
unchanged, we get
\begin{eqnarray}
P_{E}&=&\left(\frac{m_V \sqrt{1-e^2_{V,t}}}{m_E \sqrt{1-e^2_{E,t}}} 
(P^{1/3}_{V,t}-P^{1/3}_{V})+P^{1/3}_{E,t} \right)^3  \;.
\end{eqnarray}

The results are  illustrated in Figure 4b. 
What becomes immediately visible is the extreme sensitivity
of $P_{Hal}$ with respect to very minor changes of
$P_V$ and $P_E$. Indeed, the resulting orbital period 
of Venus, $P_V=0.61505$\,years, represents only 
a decrease of 0.023 per cent with respect to 
today's value $P_V=0.61519$\,years.
For the Earth, the corresponding value is 
0.015 per cent (1.00015 vs 1 year).
For the angular momenta those values amount
to 0.0077 and 0.005 per cent, respectively.

While we should, in principle, also consider changes to the sidereal year 
as the basic unit of time by which any physical period on Earth is 
measured, we now see that this very minor 
modification is essentially irrelevant to the 
interpretation of the final results.

\subsection{Robustness check with modified eccentricities}

Since the assumption of constant eccentricities as used
above is certainly not fully justified, in the following
we assess the robustness of our 
inversion against modified eccentricities.  
Although the current value of Jupiter's eccentricity is 
already close to the maximum predicted by Laplace–Lagrange 
secular perturbation theory, we nevertheless investigate what 
would happen if this value were increased to 0.1.
It might indeed be worthwhile to 
assess those high values, particularly 
in light of recent theories that attempt to explain
the late Precambrian ``snowball Earth'' in terms of 
a higher frequency of impacts from the asteroid belt
\cite{Fu2024}. These are known to be  
strongly fostered by high
eccentricities of Jupiter 
(see, e.g., Figure 1 in \citeA{Horner2012}).

Figure 5a shows the results of redoing the computations from the previous 
subsection, but with a high eccentricity of 0.1 for Jupiter and a zero 
eccentricity for Saturn, and vice versa (with this pairwise choice of 
high and low eccentricity, we acknowledge their typical anti-phase evolution; see
\citeA{Michtchenko2001} and \citeA{Ito2000}). 
For this computation, however, Equation (7) must be 
generalized to 
\begin{eqnarray}
P_{S}&=&\left( \frac{m_J \sqrt{1-e^2_{J,t}} P^{1/3}_{J,t}+ 
                     m_S \sqrt{1-e^2_{S,t}} P^{1/3}_{S,t}-
                     m_J \sqrt{1-e^2_{J}} P^{1/3}_{J}}
                     {m_S \sqrt{1-e^2_{S}}} \right)^3  \;.
\end{eqnarray}

\begin{figure}
	\includegraphics[width=0.99\textwidth]{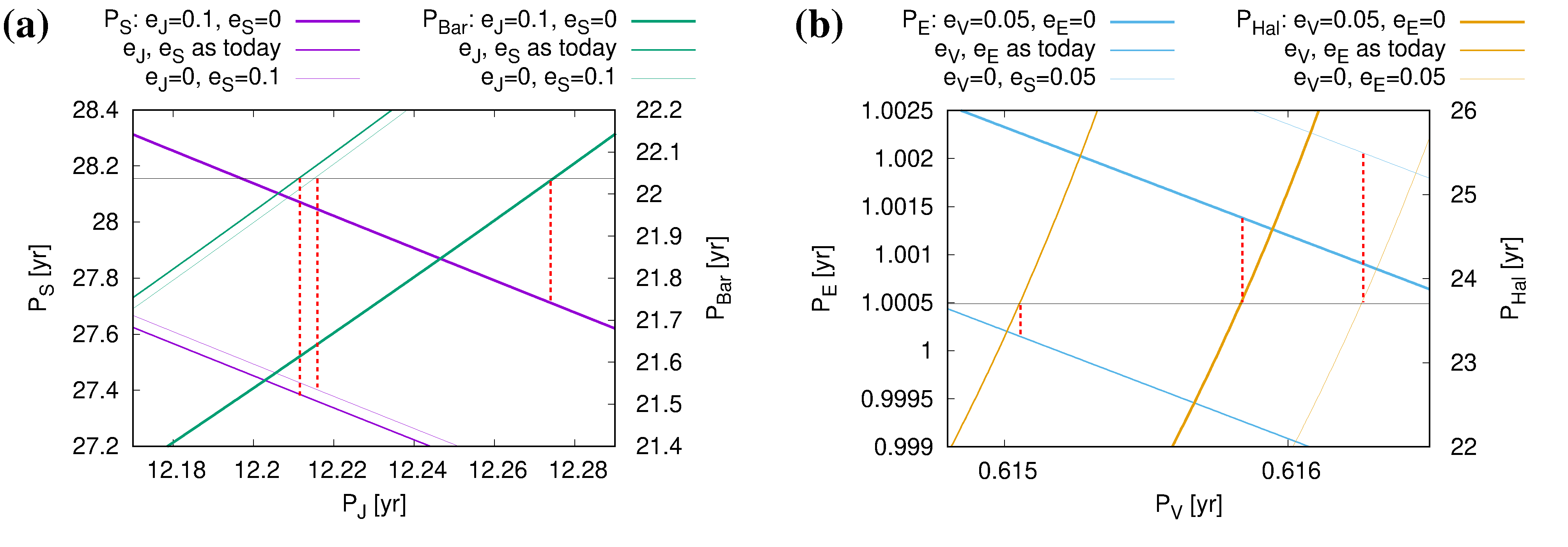}
	\caption{Graphical solution of Equation (2) assuming the 
    pairwise angular-momentum constraint of Equation (9) for 
    Jupiter and Saturn (a), and the corresponding constraint 
    of Equation (10) for Venus and Earth (b).  In addition to using 
    today's eccentricities, we also consider two cases with modified 
    eccentricities. For Jupiter and Saturn, these are 0.1 for one 
    planet and 0 for the other. For Venus and Earth, the corresponding 
    eccentricities are 0.05 and 0.}
	\label{fig:fig6}
\end{figure}

Remarkably, the solutions to our inverse problem remain quite 
robust even when eccentricities as high as 0.1 are included. Specifically,
for the choice $e_{J}=0.1$ and $e_{S}=0$ we obtain
the slightly increased values of $P_J=12.27$\,years and  $P_S=27.70$\,years,
while for $e_{J}=0$ and $e_{S}=0.1$ we obtain the nearly unchanged values
of $P_J=12.21$\,years and  $P_S=27.40$\,years.
Therefore, the derived periods appear to be relatively reliable, provided that 
the Elatina series does indeed represent a solar activity proxy and 
our synchronization theory is fundamentally correct.

Figure 5b shows the corresponding results for Venus and Earth, when 
generalizing Equation (8) to
\begin{eqnarray}
P_{E}&=&\left( \frac{m_V \sqrt{1-e^2_{V,t}} P^{1/3}_{V,t}+ 
                     m_E \sqrt{1-e^2_{E,t}} P^{1/3}_{E,t}-
                     m_V \sqrt{1-e^2_{V}} P^{1/3}_{V}}
                     {m_E \sqrt{1-e^2_{E}}} \right)^3  \;.
\end{eqnarray}
We see that even when increasing one of the excentricities
of Venus or Earth 
to a high value of 0.05 (note that Laskar's 
very long and highly accurate simulations had 
found variations between 0 and 0.06 for Earth \cite{Laskar2011}), 
the increase of $P_V$ and $P_E$
still remains below of 0.2 per cent.

\section{Conclusions}

In this paper, we have re-examined the Elatina series 
of sedimentary laminae from South Australia, which can be 
interpreted in terms of both a ``tidal theory'' and 
a ``solar theory''.
When interpreting the main period as a modified solar Hale cycle of 
23.7\,years, and the longer-term ``Elatina cycle'' of 314\,years 
as a modified Suess-de Vries cycle, we found a similar anti-correlation between the 
length of cycle $n$ and the amplitude of cycle $n+1$ as is known today.

We then focused on the phase stability of the alleged Schwabe cycle, 
which we assessed by analyzing Dicke's ratio for the residuals of the 
cycle minima. After subtracting the strong 
$\sim 314$-year cycle and its first overtone, we found a nearly 
perfect clocking within two time segments divided approximately 
at the ``varve year'' 600, where five relatively long intervals 
occurred consecutively. 
Interestingly, the replacement of the cycle minima by the
maxima before that break-point lead to a 
nearly perfect convergence of Dicke's ratio to the 
theoretical value of 0.5 for a clocked process.  While, in the 
first instance, such a replacement may seem contrived, 
it could make perfect sense in light of the recent finding 
of 90$^{\circ}$ phase jumps in numerical simulation of the solar cycle
\cite{Stefani2025}. 
It could also shed new light 
on corresponding phase jumps observed in algae data from the Holocene.

Both the period-amplitude anti-correlation, and the 
90$^{\circ}$ phase jump, provide substantial arguments for the ``solar theory'' 
of the Elatina series. As far as we are aware, the competing ``tidal theory'' 
\cite{Zahnle1987,Williams1989,Deubner1990,Williams2023}
is less suitable to explain these phenomena.

Thus motivated, and encouraged by the recent successes of our 
tidally synchronized solar dynamo model in explaining various 
solar activity cycles, we asked what planetary orbits would be 
required to also explain the Elatina series observations.
Our task was, therefore,  to explain a 7 
per cent change in the Hale cycle and a 63 
per cent change in the Suess-de Vries cycle simultaneously.
When tentatively assuming
pairwise angular momentum conservation of 
Jupiter/Saturn and Venus/Earth, respectively, our 
inverse problem turned out to be
unique when keeping the eccentricities constant. 
For Jupiter, we arrived at a 3 per cent increase in its orbital 
period, which equates to a 1 per cent increase in angular momentum.
The corresponding change in the Earth's orbital period was found to be 
0.015 per cent, which equates to a change in angular momentum of 0.005 per cent 
(similar values apply for Venus). The latter value would increase 
to 0.2 per cent if large changes to the eccentricities of the Earth and 
Venus were allowed.

The reason for the significant effect of relatively minor modifications to 
orbital periods on the Hale and Suess-de Vries cycles is that both cycles 
represent beat periods, which include small differences between large 
numbers in the respective denominators.

As non-experts in celestial mechanics, we are not in a 
position to validate our results numerically. They do not 
appear to be completely unrealistic, though.
Admittedly, the usual eccentricity swaps
between Jupiter and Saturn, as computed in frame of the
Laplace-Lagrange secular evolution theory 
\cite{Fitzpatrick2012},
imply only angular momentum changes
of around 0.15 per cent for Jupiter and 0.34 
per cent for Saturn, which are markedly below
the respective values of 1 per cent and 2.3 per cent 
as derived by us. Yet, this theory is only approximate 
in nature and capable of predicting the secular evolution 
of the solar system with reasonable accuracy for up to a million 
or so years. Over longer timescales, 
it becomes inaccurate because the true long-term dynamics of the 
solar system contains chaotic elements resulting from resonances.
As shown by \citeA{Psychoyos2004}, Jupiter-Saturn-like 
planetary pairs in 5:2 resonance
can easily develop variations of their semi-major axes
in the order of a few per cent, with that of ``Saturn''
being larger, see Figure 8 in that paper.
This applies already for typical eccentricity variation 
in the order of 0.1, while unstable orbits,
including escapes, can also develop 
for slightly different parameters.

Undoubtedly, the inferred ratio of the ancient 
periods of Saturn and Jupiter (2.24) differs markedly from the 
present value (2.48), which is close to the aforementioned 5:2 resonance.
While it might be tempting to speculate about the transition 
from the 2:1 resonance in the early phase of solar system 
evolution to the current 5:2 resonance, this could be misleading 
as that transition is typically assumed to have occurred much earlier. 
Another speculation 
is related to the impact theory \cite{Fu2024} 
of the ``snowball Earth'' 
period to which the Elatina formation belongs.
As shown by \citeA{Horner2012}, a strongly 
enhanced eccentricity
of Jupiter of 0.1 could have increased the 
impact frequency
of asteroids by a factor of 2.
It is therefore legitimate to ask then whether the
significant period change as argued for here 
could indeed be connected with a corresponding
change of the eccentricity.

In any case, this paper was not an attempt to debunk the 
alternative ``tidal theory'' of the Elatina series. Instead, it was intended 
to provide an encouraging response to the ``What if?'' question posed in the 
title.

%
%

\section*{Open Research Section}
The data for the Elatina series used for Figure 1 
were obtained by digitizing Figures
1 and 2 of \citeA{Bracewell1986}. The data for the recent sunspot cycles, as
used in Figure 2, were obtained from \citeA{Hathaway2015}.

\section*{Conflict of Interest declaration}
The authors declare there are no conflicts of interest for this manuscript.

\acknowledgments
This work received funding from the Helmholtz Association 
in frame of the AI project GEOMAGFOR (ZT-I-PF-5-200), and from Deutsche 
Forschungsgemeinschaft under the grant no. MA 10950/1-1.

%
%


\end{document}